\input{aipcheck}

\documentclass[
    ,final            
  ]
  {aipproc}

\layoutstyle{8x11double}

\begin{document}

\title{Production of strangeness -1 and -2 hypernuclei within a field theoretic model}




\author{R. Shyam}{ address={Saha Institute of Nuclear Physics, 1/AF Bidhan Nagar, Kolkata 
700064, India} }


\begin{abstract}
We present an overview of a fully covariant formulation for describing the
production of strangeness -1 and -2 hypernuclei using probes of different 
kinds. This theory is based on an effective Lagrangian picture and it focuses
on production amplitudes that are described via creation, propagation and
decay into relevant channel of intermediate baryonic resonance states in
the initial collision of the projectile with one of the target nucleons.
The bound state nucleon and hyperon wave functions are obtained by solving
the Dirac equation with appropriate scalar and vector potentials. Specific
examples are discussed for reactions which are of interest to current and
future experiments on the hypernuclear production.

\end{abstract}

\maketitle


\section{Introduction}
Hypernuclei represent the first kind of flavored nuclei (with new quantum
numbers) in the direction of other exotic nuclear systems (e.g., charmed
nuclei). They introduce a new dimension to the traditional world of atomic
nuclei. With a new degrees of freedom (the strangeness), they provide a
better opportunity to investigate the structure of atomic nuclei~\cite{gib95}.
For example, since the hyperons do not suffer from the restrictions of the 
Pauli's exclusion principle, they can occupy all the states that are
already filled up by the nucleons right upto the center of the nucleus. This
makes them unique tools to investigate the structure of the deeply bound
nuclear states (see, e.g.,~\cite{has06,ban90}). The data on the
hypernuclear spectroscopy have been used extensively to extract information
about the hyperon-nucleon ($YN$) interaction within a variety of theoretical
approaches~\cite{hiy00,kei00,gul12}.

$\Lambda$ hypernuclei [they correspond to strangeness ($S$) -1] can be produced 
by beams of mesons, protons and also heavy ions. The electromagnetic probes like 
photons and electrons can also be used for this purpose. Although, the stopped 
as well as in-flight $(K^-,\pi^-)$~\cite{ban90,chr89} and $(\pi^+,K^+)$
\cite{has06,ban90,chr89} reactions have been most extensively used in the 
experimental investigations of their production, the feasibility of producing 
them via the $(p,K^+)$~\cite{kin98,shy04,shy06}, $(\gamma,K^+)$
\cite{yam95,shy08,shy09} and $(e,e^\prime K^+)$
~\cite{miy03,iod07,yua06,cus09,ach12,nak13} reactions has also been demonstrated 
in the recent years. The $(K^-,K^+)$ reaction is one of
the most promising ways of studying the $S = -2$ $\Xi$ hypernuclei becuase it 
leads to the transfer of two units of both charge and strangeness to the target 
nucleus.  There is an experimental proposal approved at JPARC facility in Japan 
to produce such nuclei using this reaction.

Several features of various hypernuclear production reactions can be understood 
by looking at the corresponding momentum transfers to the recoiling nucleus 
becuase it controls to some extent the population of the hypernuclear states. 
In Fig.~1, the momentum transferred to the recoiled nucleus is shown as a 
function of beam energy at two angles of the outgoing kaon for a number of 
reactions. We see that the $(K^-,\pi^-)$ reaction allows only a small momentum 
transfer to the nucleus (at forward angles), thus there is a large probability 
of populating $\Lambda$-substitutional states in the residual hypernucleus
($\Lambda$ occupies the same angular momentum state as that of the replaced 
neutron). On the other hand, in $(\pi^+,K^+)$, $(\gamma,K^+)$ and $(K^-,K^+)$ 
reactions the momentum transfers are larger than the nuclear Fermi momentum. 
Therefore, these reactions can populate hyperon states with configurations of 
a nucleon hole and a hyperon in a series of orbits covering all bound states. 
The momentum transfers involved in the $(p,K^+)$ reaction are still larger by 
a factor of about 3. Thus, the states of the hypernuclei excited in the 
$(p,K^+)$ reaction may have a different type of configuration as compared to 
those excited, e.g., in the $(\pi^+,K^+)$ reaction. Nevertheless, it should be
mentioned that usually larger momentum transfers are associated with smaller 
hypernuclear production cross sections. Each reaction has its own advantage and 
plays its own role in a complete understanding of the hypernuclear spectroscopy.

In this paper, we present a short review of the investigations made by us on the
production of $S$ -1 and -2 hypernuclei within an effective Lagrangian model
\cite{shy04,shy06,shy08,shy09,ben10,shy12}. It retains the full field theoretic 
structure of the interaction vertices and treats baryons as Dirac particles. 
The initial state interaction of the incoming projectile with a free or bound 
target nucleon leads to excitation of intermediate baryon or hyperon resonant 
states, which propagate and subsequently decay into a $K^+$ meson and hyperon
that gets captured into one of the nuclear orbits to produce the hypernucleus. 
 
\begin{figure}
\includegraphics[width=\columnwidth]{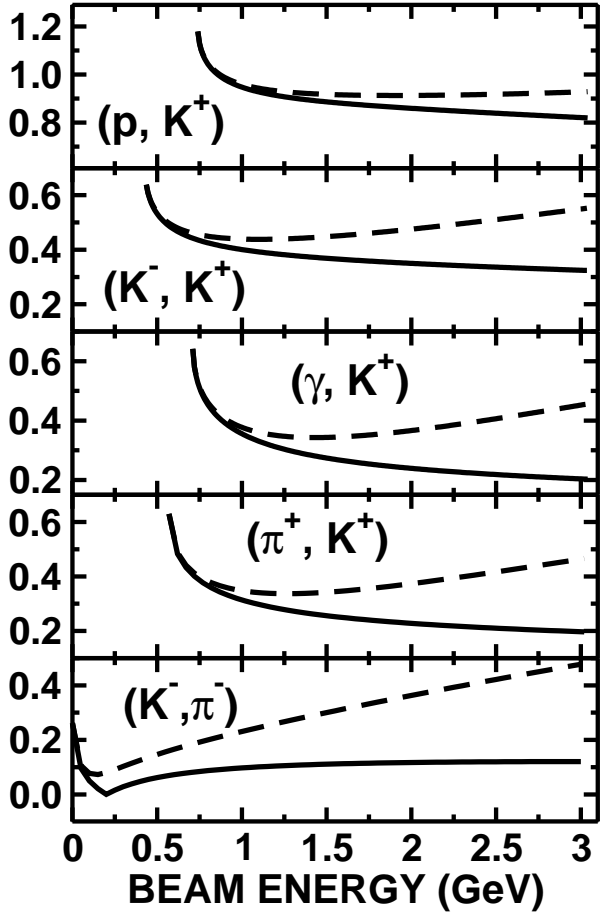}
\caption{\label{Fig.1} 
The momentum transfer involved in various hypernuclear production
reactions as function of beam energy for the outgoing kaon angles of
$0^\circ$ (full lines) and $10^\circ$ (dashed lines) for the $^{12}$C
target.}
\end{figure}

\section{Covariant hypernuclear production amplitudes}

Since we are still far way from calculationing the intermediate energy
scattering and reactions directly from the lattice QCD, the effective field
theoretical description in terms of the baryonic and mesonic degrees
of freedom is usually employed to describe these processes. These approaches
introduce the baryonic resonance states explicitly in their framework and
QCD is assumed to provide justification for the parameters or the
cut-off functions used in calculations.
\begin{figure}
\includegraphics[width=\columnwidth]{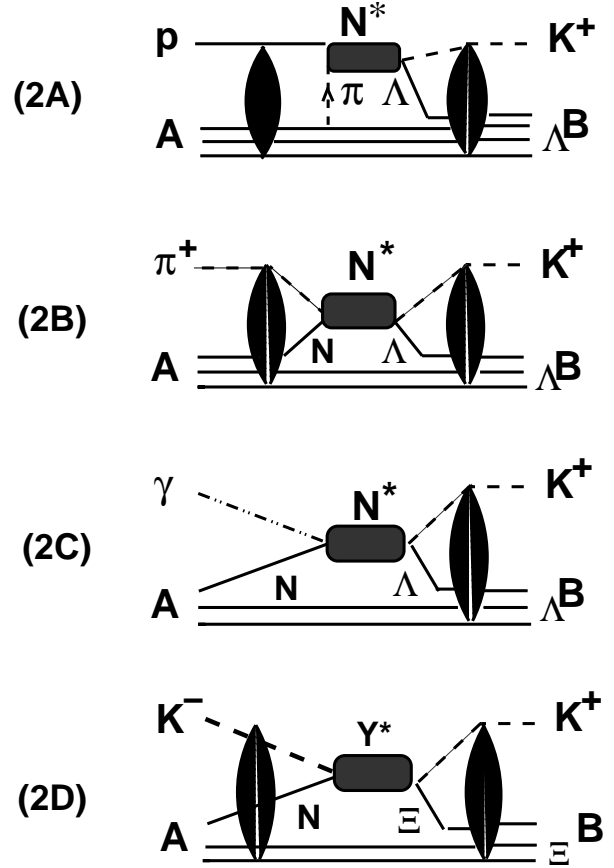}
\caption{\label{Fig.2}
Types of the Feynman diagrams included in calculations of various reactions. 
In case of the $(p,K^+)$ reaction another diagram also contributes to the 
total amplitude. This is known as the projectile emission diagram, where the
exchanged pion emerges from the projectile vertex. The elliptic shaded areas 
represent the optical model interactions in the incoming or outgoing channels.}
\end{figure}

We use the diagrams shown in Fig.~2 for the calculations of the hypernuclear 
production reactions. In all the cases, the initial state interaction of the 
projectile with a bound nucleon of the target leads to excitations of intermediate 
resonance intermediate states that decay into kaon and the hyperon. The latter gets 
captured into one of the nuclear orbits. In case of reactions leading to the 
production of the $\Lambda$ hypernuclei, we have included three nucleon resonances,
$N^*$(1650)[$\frac{1}{2}^-$], $N^*$(1710)[$\frac{1}{2}^+$] $N^*$(1720)
[$\frac{3}{2}^+$], becuase they have appreciable branching rations for decay into 
the $K^+\Lambda$ channel and are known to contribute predominantly to the 
corresponding elementary reactions involved in various processes
\cite{shy09,shy99,shk05}. On the other hand, in case of the $(K^-,K^+)$ reaction 
leading to the production of $\Xi$ hypernuclei, the intermediate channels are the 
hyperon states.  We have considered the $\Lambda$, $\Sigma$ hyperons and eight of 
their resonances with masses up to 2.0 GeV [$\Lambda(1405)$, $\Lambda(1520)$, 
$\Lambda(1670)$, $\Lambda(1810)$, $\Lambda(1890)$, $\Sigma(1385)$, $\Sigma(1670)$, 
$\Sigma(1750)$], which are represented by $Y^*$ in Fig.~2 (2D). These resonances 
have also been considered in a description of the $\Xi$ hyperon production in the 
elementary $p(K^-,K^+)\Xi^-$ reaction within a similar model where a good 
description of the available total and differential cross sections has been 
obtained~\cite{shy11}.  

To calculate the amplitudes corresponding to the diagrams shown in Fig.~2, 
one requires the effective Lagrangians at the meson-baryon-resonance vertices 
(which involve coupling constants and the form factors), and the propagators 
for various resonances. They have been taken to be the same as those given Refs.
\cite{shy06,shy09,shy11,shy12}. In addition, one needs spinors for the nucleon 
hole and hyperon particle bound states. They are obtained by following the 
procedure as discussed in the next section. In our model terms corresponding to 
the interference between various components of a given diangram are retained in 
the corresponding total production amplitude. Since calculations within this 
theory are carried out in the momentum space all along, they includes all the 
nonlocalities in the production amplitude that arises from the resonance propagators.

\section{Results and discussions}

\subsection{1. Hyperon bound state spinors}

The spinors for the final hypernuclear bound state and for intermediate
nucleonic states that are required for the calculations of various amplitudes,
have been calculated within a phenomenological model where we assume these 
states to have pure-single particle or single-hole configurations. The hyperon 
bound states have been calculated in a phenomenological model where they are 
obtained by solving the Dirac equation with scalar and vector fields having a 
Woods-Saxon (WS) radial form. With a set of radius and diffuseness parameters, 
the depths of these fields are searched to reproduce the binding energy (BE) 
of a given state. Since the experimental values of the BEs for the $\Xi^-$ bound 
states are as yet unknown, we have adopted the predictions of the latest version 
of the quark-meson coupling (QMC) model~\cite{gui08} for the corresponding BEs in 
our search procedure for these states. The spinors for the nucleon-hole states 
were also calculated in the same way. For a comprehensive discussion of these 
calculations we refer to, e.g.,~\cite{shy12,kaz12}. In studies of the 
$(\gamma,K^+)$ and $(K^-,K^+)$ reactions in ref.~\cite{shy09} and \cite{shy12},
respectively, the spinors calculated within the latest version of the QMC model
have also been used. The cross sections calculated by using these spinors are
similar to those obtained with the phenomenological model 
The spinors in the momentum space are obtained by Fourier transformation of the 
corresponding coordinate space spinors. 
\begin{figure}
\includegraphics[width=\columnwidth]{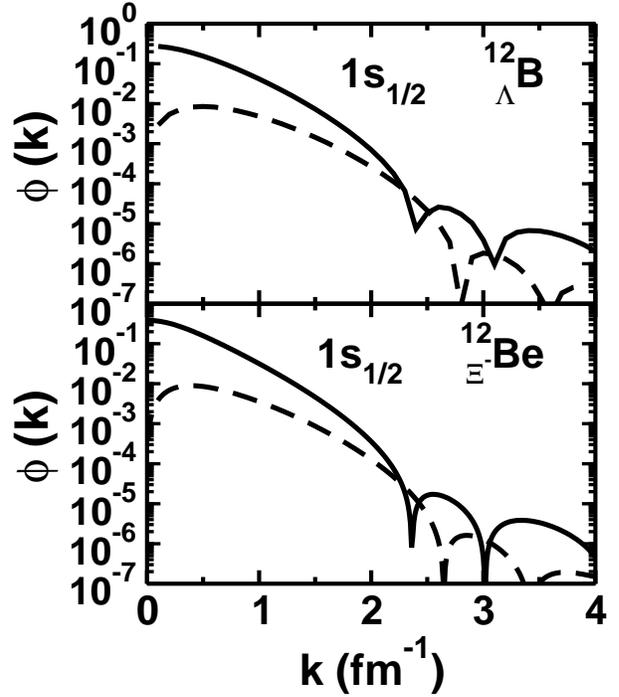}
\caption{\label{Fig.3}
Magnitudes of the upper (full line) and lower (dashed line) components of the
momentum space spinors for the hyperon orbit as shown.}
\end{figure}

In Fig.~3, we show the lower and upper components of the Dirac spinors in momentum 
space for the  $0s_{1/2}$ $\Lambda$ and $\Xi$ hyperons in $^{12}\!\!\!_\Lambda$B 
and $^{12}{\!\!\!_{\Xi^-}}$Be hypernuclei that are generated, respectively, via the 
$(\gamma,K^+)$ and $(K^-,K^+)$ reactions on a $^{12}$C target. The experimental 
BE of the $\Lambda$ state is 11.37 MeV (see, e.g., Refs.~\cite{miy03,yua06,iod07}).
Since the experimental values of the BE for the $\Xi^-$ bound state is still unknown, 
we have adopted in our search procedure the corresponding value predicted in the latest 
version of the quark-meson coupling (QMC) model~\cite{gui08} which is 3.04 MeV
\cite{kaz12}. With a radius and diffuseness parameters of 0.983 fm and 0.606 fm,
the searched values of $V_s$ and $V_v$ are -212.70 MeV and 171.78 MeV, and -112.11
MeV and 90.81 meV for the $\Lambda$ and $\Xi$ states, respectively. The values of 
$V_s$+ $V_v$ that should be comparable to the depth of the corresponding
conventional Woods-Saxon potentials, are -40.92 Mev and -21.30 MeV for $\Lambda$ and
$\Xi$ interactions, respectively. Thus depth of the $\Xi$-nucleus interaction 
($V_{\Xi}$) is  about half of the $\Lambda$-nucleus interaction ($V_\Lambda$). 
Nevertheless, our $V_{\Xi}$ is towards the higher side of the values available in
the literature. While our potential is closer to that deduced in Ref.~\cite{dov83} 
(although their radius parameter was slightly bigger than ours), it is higher than 
the shallow potential ($V_{\Xi}$ $\approx$ -16 MeV) conjectured from the emulsion 
events in the E176 experiment~\cite{aok93}.   

In each case, we note that only for momenta $<$ 1.5 fm$^{-1}$, is the lower 
component of the spinor substantially smaller than the upper component. In the 
region of momentum transfer pertinent to the reactions under study, the lower 
components of the spinors are not negligible as compared to the upper component. 
This clearly demonstrates that a fully relativistic approach is essential for
an accurate description of this reaction. In each case the momentum density 
of the hyperon shell, in the momentum region around 0.35 GeV/c, is at least 
2-3 orders of magnitude larger than that around 1.0 GeV/c. Thus reactions 
involving lower momentum transfers are expected to have larger cross sections.

\subsection{2. Production of Strangeness -1  hypernuclei}

The $\Lambda$ hypernuclei are the most studied $S$ = -1 hypernuclear systems. 
They have been investigated using both hadronic (e.g., $K^-$, $\pi$) and 
electromagnetic (photon and electron) probes~\cite{has06}. A comparison of 
relative merits of various probes of the production of $\Lambda$ 
hypernuclei may be useful here. In contrast to the hadronic reactions e.g.  
$[ (K^-,\pi^-)$ and $(\pi^+,K^+)]$, which take place mostly at the nuclear 
surface due to strong absorption of both $K^-$ and $\pi^\pm$, the 
$(\gamma,K^+)$ and $(e,e^\prime K^+)$ reactions occur deep in the nuclear 
interior since $K^+$-nucleus interaction is weaker. Thus, this reaction is 
an ideal tool for studying the deeply bound hypernuclear states if the 
corresponding production mechanism is reasonably well understood. While 
hadronic reactions excite predominantly the natural parity hypernuclear 
states, both unnatural and natural parity states are excited with comparable 
strengths in the electromagnetic reactions~\cite{ben89,mot94,lee95,lee01}.
This is due to the fact that sizable spin-flip amplitudes are present
in the elementary photo-kaon production reaction, $p(\gamma,K^+)\Lambda$,
since the photon has spin 1. This feature persists in the hypernuclear
photo- and electro-production. Furthermore, since in these reactions a
proton in the target nucleus is converted into a hyperon, it leads to the
production of neutron rich hypernuclei (see, e.g., Ref.~\cite{agn12}),
which may carry exotic features such as a halo structure. It can produce
many mirror hypernuclear systems which would enable the study of the charge 
symmetry breaking with strangeness degrees of freedom. Our effective Lagrangian 
model has been used for describing the hypernuclear production using both the 
hadronic $(p,K^+)$~\cite{shy04,shy06} and $(\pi^+,K^+)$~\cite{ben10}, and the 
electromagnetic $(\gamma,K^+)$~\cite{shy09} reactions on a number of target 
nuclei. In the following we present some highlights of our calculations for 
the $\Lambda$ hypernuclear production using the $(\gamma,K^+)$ reaction.

\subsection{2.1 $\Lambda$ hypernuclear production via $(\gamma,K^+)$ reaction} 
  
The threshold for the $^{12}$C($\gamma,K^+$)$^{12}\!\!\!_\Lambda$B reaction is
about 695 MeV. The momentum transfer involve in this reaction at 10$^\circ$ kaon 
angles varies between approximately 2 fm$^{-1}$ to 1.4 fm$^{-1}$ for photon 
energies ranging between 0.7 GeV to 1.2 GeV~\cite{shy08}. In Fig.~4, we show 
the differential cross section at $K^+$ angle of 10$^\circ$  as a function of 
the photon energy (in the range 0.7-1.2 GeV) for this reaction. The hypernuclear 
states populated are $1^-$, $2^-$, and $2^+$, $3^+$ corresponding to the  
particle-hole configurations of $(1p_{3/2}^{-p},1s_{1/2}^\Lambda)$ and 
($(1p_{3/2}^{-p},1p_{3/2}^\Lambda)$, respectively. We  note that within 
each configuration the highest $J$ state is most strongly excited. Furthermore, 
unnatural parity states within each group are preferentially excited by this 
reaction. The unnatural parity states are excited through the spin flip process. 
Thus this confirms that kaon photo- and also electro-production reactions on 
nuclei are ideal tools for investigating the structure of unnatural parity 
hypernuclear states. The addition of unnatural parity states to the spectrum 
of hypernuclei is expected to constrain the spin dependent part of the
effective $\Lambda-N$ interaction more tightly.

Another noteworthy aspect of Fig.~4 is that cross sections peak at photon
energies around 850 MeV, which is about 200 MeV above the production
threshold for this reaction. Interestingly, the total cross section of
the elementary $p(\gamma,K^+)\Lambda$ reaction also peaks about the same
energy above the corresponding production threshold (~910 MeV). Therefore,
the $(\gamma,K^+)$ reaction on nuclei has the same basic features as that
of the corresponding elementary reaction.
\begin{figure}
\includegraphics[width=\columnwidth]{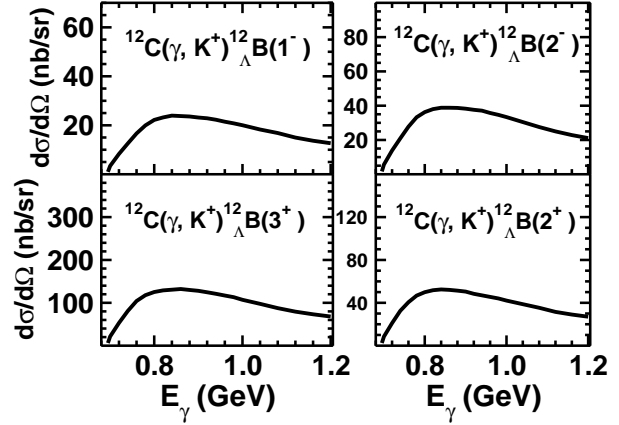}
\caption{\label{Fig.4}
Differential cross sections (for the outgoing kaon angle of 10$^\circ$) for 
the $^{12}$C$(p,K^+)$$^{12}\!\!\!_\Lambda$B reaction leading to hypernuclear 
states as indicated.}
\end{figure}

\subsection{3. Production of Strageness -2 hypernuclei}

The $(K^-,K^+)$ reaction provides one of the most promising ways of studying the 
$S = -2$ hypernuclei.  This reaction implants a $\Xi$ hyperon in the nucleus 
through the elementary $p(K^-, K^+)\Xi^-$ process. Therefore, it is essential to 
investigate first the $(K^-,K^+)$ reaction on a proton target leading to the 
production of a free $\Xi^-$ hyperon. This has been done in Ref.~\cite{shy11}
within a similar effective Lagrangian model, where a good agreement has been 
achieved with available with experimental total and differential cross sections.
The input information extracted from this study is then used in the description 
of the formation of $\Xi^-$-hypernuclei using this reaction.

\subsection{3.1 Production of $\Xi$ hypernuclei via $(K^-,K^+)$ reaction} 

In this section, we discuss strangeness -2 hypernuclear production reactions
$^{12}$C$(K^-,K^+)^{12}{\!\!\!_{\Xi^-}}$Be and $^{28}$Si$(K^-,K^+)^{28}
{\!\!\!_{\Xi^-}}$Mg. The threshold beam momenta for these reactions are about 0.761 
GeV/c and 0.750 GeV/c, respectively. We have employed pure single-particle-single-hole 
$(p^{-1}\Xi)$ configurations to describe the nuclear structure part. The QMC model 
predicts only one bound state for the $^{12}{\!\!\!_{\Xi^-}}$Be hypernucleus with a  
binding energy of 3.038 MeV and quantum numbers ($1s_{1/2}$). On the other hand, 
for $^{28}{\!\!\!_{\Xi^-}}$Mg it predicts 3 distinct bound $\Xi^-$ states, $1s_{1/2}$, 
$1p_{3/2}$, $1p_{1/2}$, with corresponding binding energies of 8.982 MeV, 4.079 MeV, 
and 4.414 MeV, respectively~\cite{kaz12}. These binding energies were used in our 
search procedure to obtain the spinors of the corresponding bound states.

In case of the $^{12}$C target, the $\Xi^-$ hyperon in a 1$s_{1/2}$ state can
populate 1$^-$ and 2$^-$ states of the hypernucleus corresponding to the
particle-hole configuration $[(1p_{3/2})^{-1}_p,(1s_{1/2})_{\Xi^-}$]. The states
populated for the $^{28}{\!\!\!_{\Xi^-}}$Mg hypernucleus are [2$^+$, 3$^+$],
[1$^-$, 2$^-$, 3$^-$, 4$^-$], and [$2^-$, $3^-$] corresponding to the
configurations $[(1d_{5/2})^{-1}_p, (1s_{1/2})_{\Xi^-}$], $[(1d_{5/2})^{-1}_p, 
(1p_{3/2})_{\Xi^-}$], and $[(1d_{5/2})^{-1}_p, (1p_{1/2})_{\Xi^-}$], respectively.
In Fig.~5, we have shown results for populating the hypernuclear state with
maximum spin of natural parity for each configuration.  We have used a plane
wave approximation to describe the relative motion of kaons in the incoming and
outgoing channels. However, distortion effects are partially accounted for by
introducing reduction factors to the cross sections as described in Ref.
\cite{ike94}. 

It is clear from Fig.~5 that for both the hypernuclear production reactions, the
cross sections peak at $p_{K^-}$ around 1.0 GeV/c which is $\approx$ 0.3 GeV/c 
above the production thresholds of the two reactions. It is not too different from 
the case of the elementary $\Xi^-$ production cross sections where the peaks of the
cross sections occur at about 0.35 -0.40 GeV/c above the corresponding production
threshold (see Ref.~\cite{shy11}). 

The magnitudes of the cross sections for the $^{12}{\!\!\!_{\Xi^-}}$Be production 
are in excess of 1 $\mu b$ near the peak position.  It is important to note that at 
the beam momentum of 1.6 GeV/c, the magnitude of our cross section for this case is 
similar to that obtained in Ref.~\cite{ike94} within an impulse approximation model. 
Moreover, our cross sections at 1.8 GeV/c also are very close to those of Ref.
\cite{dov83} for both the targets. A more rigorous consideration of the distortion 
effects could alter the pattern of the beam momentum dependence, {\it e.g.} it is 
likely to be relatively stronger at lower values of $p_{K^-}$ as compared to higher 
values.  These effects will be investigated in a future publication.    

\begin{figure}
\includegraphics[width=\columnwidth]{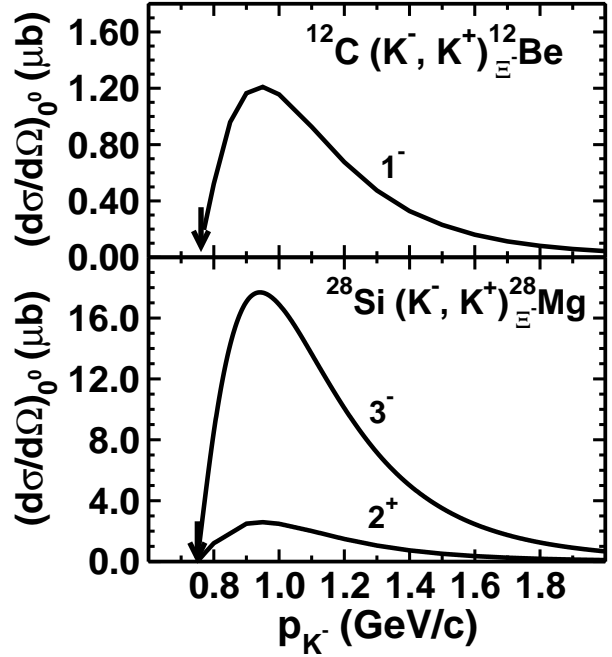}
\caption{\label{Fig.5}
Differential cross section at $0^\circ$ as a function of
$K^-$ beam momentum for the $^{12}$C$(K^-,K^+)^{12}{\!\!\!_{\Xi^-}}$Be
and $^{28}$Si$(K^-,K^+)^{28}{\!\!\!_{\Xi^-}}$Mg reactions. The spin-parity
of the final hypernuclear states are indicated on each curve. 
}
\end{figure}
\section{4. Summary and conclusions}

In summary, it is clear that hypernuclear spectroscopy is indispensable for 
the quantitative understanding of the strangeness -1 and -2 hypernuclear 
structure and the hyperon-nucleon interaction. Driven by the new electron 
and kaon accelerator facilities, the field of hypernuclear production 
reactions is expected to experience a big boost. Already some data on the 
electron induced reactions leading to the formation of $\Lambda$ hypernuclei 
are available from the Jefferson Laboratory and more such data are expected 
to be available from the MAMI-C facility in the near future. Moreover, data 
on the strangeness -2 $\Xi$ hypernucei will soon be available from the JPARC 
facility in Japan. At the same time, our understanding of the hypernuclear 
production mechanism has improved significantly over the last decade.

A fully relativistic approach is essential for an accurate description of
the hypernuclear production cross sections. It is feasible to calculate
the reactions induced by  hadronic and electromagnetic probes within a
single fully covariant effective Lagrangian picture. Since the relevant
elementary production cross sections are also described within the similar 
picture, most of the input parameters needed for the calculations of the 
hypernuclear production are fixed independently.

The feasibility of one such model has been investigated by us. It provides 
a good account of the available data on hypernuclear production via the 
$(\pi^+,K^+)$ reaction on a number of target nuclei. It describes well the 
imporatant features of the $\Lambda$ and $\Xi$ hypernuclear production cross 
sections via $(\gamma,K^+)$ and $(K^-,K^+)$ reactions, respectively. An important
feature of our calculation is that in both the cases the production cross
sections peak at projectile ($\gamma$ or $K^-$) energies that are above the 
corresponding thresholds by almost the same amount as are the positions of the 
maxima away from their thresholds in the relevant elementary cross sections.   

Some of the predictions of our model for the strangeness -2 hypernuclear 
production should be tested soon by experiments to be performed at the JPARC
facility in Japan.

\begin{theacknowledgments} 
The author wishes to acknowledge useful discussions with H. Lenske, U. mosel 
O. Scholten, A. W.  Thomas, K. Tsushima.
\end{theacknowledgments}



\bibliographystyle{aipproc}   

\bibliography{sample}

\IfFileExists{\jobname.bbl}{}
 {\typeout{}
  \typeout{******************************************}
  \typeout{** Please run "bibtex \jobname" to optain}
  \typeout{** the bibliography and then re-run LaTeX}
  \typeout{** twice to fix the references!}
  \typeout{******************************************}
  \typeout{}
 }


\end{document}